\documentclass[twocolumn,superscriptaddress,showpacs,floatfix,preprintnumbers, nofootinbib,hyperref]{revtex4} 
\usepackage{graphicx}
\usepackage{rotating}
\usepackage{epsfig}
\usepackage{bm}
\usepackage[usenames,dvipsnames]{xcolor}

\usepackage{color}

\def\he4{$^4$He}
\def\h2{$^2$H}

\newcommand{\lesssim}{\,\rlap{\lower3.7pt\hbox{$\mathchar\sim$}}
\raise1pt\hbox{$<$}\,}

\begin{document}

\title{Breaking the symmetries  in     self-induced flavor conversions
\\ of neutrino beams  from a ring}

\author{Alessandro Mirizzi} 
\affiliation{Dipartimento Interateneo di Fisica ``Michelangelo Merlin'', Via Amendola 173, 70126 Bari, Italy}
\affiliation{Istituto Nazionale di Fisica Nucleare - Sezione di Bari, Via Amendola 173, 70126 Bari, Italy}


\begin{abstract}
Self-induced flavor conversions of  supernova (SN) neutrinos have been characterized in the spherically symmetric ``bulb'' model, reducing the
neutrino evolution to a one dimensional
problem along a radial direction. 
We lift this assumption,  presenting a  two-dimensional toy-model 
where neutrino beams are launched in many different directions from a ring. 
 We find that
self-interacting neutrinos  spontaneously  break  the spatial symmetries of this  model.
As a result  the flavor content and the lepton number  of the neutrino gas would acquire seizable direction-dependent variations, breaking the coherent behavior found in the
 symmetric case.
This finding would suggest that the previous results of the self-induced flavor evolution obtained in one-dimensional models should be critically re-examined.
\end{abstract}


\maketitle

\section{Introduction}

Dense neutrino gases in early universe or emitted from core-collapse supernovae (SNe) represent
unique cases to probe the effect of the neutrino-neutrino interactions on the
flavor conversions. Indeed, in these environments the neutrino-neutrino interactions would
generate a large neutrino potential $\mu \sim \sqrt{2} G_F n_\nu$ that in some cases can exceed the 
ordinary matter term $\lambda= \sqrt{2} G_F n_e$  and the neutrino vacuum oscillation frequency $\omega= \Delta m^2/2E$. 
When this situation is encountered the neutrino-neutrino potential would dominate the  flavor evolution 
producing large self-induced flavor conversions (see~\cite{Duan:2010bg} for a review). 
A vivid activity on these effects in the context of SN neutrinos has flourished since a 
decade~\cite{Duan:2005cp,Duan:2006an,Hannestad:2006nj,Fogli:2007bk}. 
Indeed, it has been realized that in the deepest SN regions self-induced effects can produce
collective neutrino oscillations, leading to peculiar spectral features in the oscillated neutrino spectra,
dubbed as spectral swaps and splits~\cite{Raffelt:2007cb,Duan:2007bt,Dasgupta:2009mg,Friedland:2010sc,Dasgupta:2010cd}.

The development of the self-induced flavor conversions is associated with \emph{instabilities}
in the flavor space that are triggered by the interacting neutrinos.
The first one to be noticed was the \emph{bimodal} instability present   even in an homogeneous
and isotropic neutrino gas~\cite{Hannestad:2006nj}. In particular, it was shown that  an ensemble initially composed of equal densities
of $\nu_e$ and $\bar\nu_e$ in the presence of a dominant neutrino-neutrino interaction term 
would exhibit in inverted mass hierarchy ($\Delta m^2 <0$) large pair-conversions of the type $\nu_e \bar\nu_e \leftrightarrow \nu_x \bar\nu_x$
even with a small mixing angle. This behavior has been explained in terms of an unstable pendulum in flavor space, 
where the instability is associated with the tiny mixing angle~\cite{Hannestad:2006nj,Duan:2007fw}.
Furthermore, it has been shown that if one introduces an anisotropy in the neutrino gas, this can dramatically change the
previous solution. Indeed, in a  non-isotropic neutrino ensemble, 
the neutrino-neutrino interaction term 
contains {\it multi-angle} effects since
the current-current nature of the low-energy weak interactions introduces
an angle dependent term  $(1-{\bf v}_{\bf p} \cdot {\bf v}_{\bf q})$ between two interacting neutrino modes~\cite{Qian:1994wh,Duan:2006an}.
In the case of a gas completely symmetric in flavor content of $\nu$ and $\bar\nu$, even a small deviation from a perfect isotropy
is enough to produce a multi-angle {\it  dechoerence} leading to a flavor equilibrium among the different neutrino species in both the mass 
hierarchies~\cite{Raffelt:2007yz}.  
Multi-angle effects have been extensively studied in the context of flavor evolution of SN
 neutrinos~\cite{Mirizzi:2010uz}, whose emission is far from isotropic.
It has been realized that
in some cases they can destroy     the collective behavior of 
the flavor evolution observed in an isotropic environment~\cite{Raffelt:2007yz,EstebanPretel:2007ec,Sawyer:2008zs}. 
Multi-angle effects can also lead to a trajectory-dependent matter term, which if  strong enough  suppresses the self-induced
conversions~\cite{Chakraborty:2011nf,Chakraborty:2011gd,Saviano:2012yh,Sarikas:2011am}. 
 In the context of SN neutrinos it has been often assumed  an axially symmetric neutrino emission in oder to
integrate out the azimuthal angle in the multi-angle kernel. However, it has been found that lifting this assumption,  neutrino-neutrino interactions can  
break axial symmetry and lead to azimuthal-angle dependent flavor 
conversions~\cite{Raffelt:2013rqa,Raffelt:2013isa,Duan:2013kba,Mirizzi:2013rla,Mirizzi:2013wda,Chakraborty:2014nma,Chakraborty:2014lsa}. 

The lesson that has been gained from these situations is that self-interacting neutrinos can 
\emph {spontaneously break}   the symmetries of the initial conditions, since small deviations from them can be dramatically amplified during the
further flavor evolution. 
This insight has recently stimulated  doubts  about the validity of the solution of the SN neutrino equations
of motion worked out   in the
 so-called  
\emph{``bulb model''}~\cite{Duan:2006an,Fogli:2007bk,EstebanPretel:2007ec}.
 In this framework  it is assumed the spherical symmetry about the center of the SN and the axial
symmetry about any radial direction. 
These two symmetries  allow one to reduce the problem to a one-dimensional evolution 
along a radial direction.   Remarkably, removing the assumption of  spherical symmetry it would necessary 
to solve   a challenging multi-dimensional problem to characterize the neutrino flavor evolution.

In this context,
in order to show how deviations  from the spatial symmetries of a system would affect the flavor evolution 
 a  simple two-dimensional  model has been recently proposed in~\cite{Duan:2014gfa}. 
 Namely,
 monochromatic neutrinos streaming in a stationary way in two directions (``left'' $L$ and ``right'' $R$, respectively)
 from an infinite boundary plane at $z=0$ with periodic conditions on $x$ and translation invariance along the $y$ direction.
Remarkably, there is a correspondence between the symmetries of the bulb model and the ones of this planar case.
 Indeed, the translational symmetry in the $x$ direction in the planar model corresponds to the spherical symmetry of the bulb-model 
and the $L$-$R$ symmetry is equivalent to the axial symmetry in the spherical case.
  By means of  a stability analysis of  the linearized equations of motion,
it has been shown in the planar model that if one perturbs 
the initial symmetries of the flavor content in both the two emission modes and along the boundary in the $x$ direction, 
 then self-induced oscillations can spontaneously break both these spatial symmetries~\cite{Duan:2014gfa}.
 In~\cite{Mirizzi:2015fva} we have recently performed a numerical study of the flavor evolution for this case.
We found that the initial small perturbations are amplified by neutrino interactions, leading to  non-trivial two-dimensional structures
 in the flavor content and lepton number of the neutrino enseble,  that would exhibit large space fluctuations. 
 
 The purpose of this paper is to develop a two-dimensional  model 
to capture more closely some of the features of the SN environment. In particular, with respect to the planar model
considered in~\cite{Duan:2014gfa,Mirizzi:2015fva} we make the following improvements:  
\emph{(i)} $\nu$ emission from  a ring mimicking the neutrino-sphere, \emph{(ii)} 
parameters inspired by the SN neutrino emissivity, \emph{(iii)}  declining neutrino density  from the boundary,
\emph{(iv)}  multi-angle effects. 
We also assume that self-induced flavor conversions would develop without any hindrance caused  by a 
large matter term.
Perturbing the neutrino emission in the translational symmetry on the ring 
and in the emission directions, we find the spontaneous breaking 
of  these  symmetries  in both normal and inverted mass hierarchies.
As a consequence the flavor content and the lepton number of the neutrino ensemble acquires seizable variations
along different lines of sight.
These findings are presented as follows.
In Sec.~II we describe the features of our two-dimensional model. We discuss the equations of motion
to characterize the two-dimensional flavor evolution. We show how  it is possible  
to solve this  problem by Fourier transforming these equations,   obtaining
a tower of ordinary  differential equations for the different Fourier modes. 
In Sec.~III we present the numerical results of our study.  We show how the breaking of the spatial symmetries produce direction-dependent variations in the flavor content of the ensemble.
Finally in Sec.~IV we discuss about future developments and we conclude.

\section{Two-dimensional model}

\subsection{Equations of motion}
Characterizing the SN neutrino flavor dynamics
amounts to follow the spatial evolution of the neutrino fluxes. For a stationary
neutrino emission,
the  Equations of Motion (EoMs) of 
the  $\nu$ space-dependent occupation numbers  ${\varrho} ({\bf r}, {\bf p})$
with momentum ${\bf p}$ at position ${\bf r}$ are~\cite{Sigl:1992fn,Strack:2005ux} 
\begin{eqnarray}
&& {\bf v} \cdot \nabla_{\bf r}\, {\varrho}
 = - i [{\sf\Omega}, \varrho]
\,\ ,
\label{eq:eom}
\end{eqnarray}
where we indicate with sans-serif vectors in flavor space, while for the ones in real space we use the bold-face. 
At the left-hand-side of Eq.~(\ref{eq:eom})  there is the Liouville operator representing the drift term proportional to the neutrino velocity
${\bf v}$, due to particle free streaming. Note that we are
 neglecting external forces
and an explicit time dependence of the occupation numbers.
On the right-hand-side of Eq.~(\ref{eq:eom}) the matrix $\Omega$ is the full Hamiltonian 
that reads
\begin{equation}
{\sf \Omega} = \frac{{\sf M}^2}{2 E} + \sqrt{2} G_F \left[{\sf N}_l + \int_{-\infty}^{+\infty} d E^{\prime}  {E^{\prime}}^2
\int \frac{d {\bf v}^\prime}{(2 \pi)^3} \varrho^{\prime} (1- {\bf v} \cdot {\bf v}^{\prime})  \right] \,\ ,
\end{equation}
where ${\sf M}^2$ is the matrix of the mass-squared, responsible of the vacuum oscillations.
The ordinary matter effects on neutrino flavor conversions is accounted by the matrix of charged lepton densities
$N_l$. Finally, the neutrino-neutrino interaction potential is represented by the last term of the right-hand-side, 
where the integral in ${d {\bf v}^\prime}$ is on the unit sphere and the occupation numbers $ \varrho^{\prime}$ depend
on ${\bf r}, E^{\prime}, {\bf v}^\prime$. Note that we use negative $E$ to denote anti-neutrinos.

\begin{figure}[!t]\centering
\includegraphics[angle=0,width=1.\columnwidth]{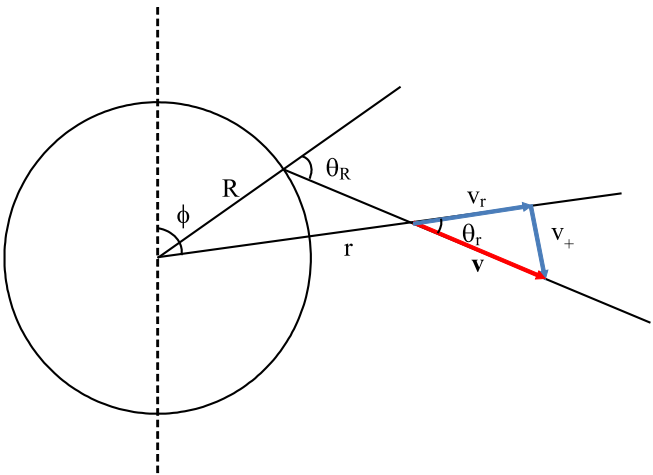}

\caption{Two-dimensional model for the neutrino beams emitted from a ring with radius $r=R$.
\label{bulb}}
\end{figure}

In order to show the effect of spontaneous  breaking of spatial symmetries we consider them to be emitted in a plane from  a ring with radius $r=R$.
We have then a two-dimensional model for which 
it is natural to use a system of polar coordinates to describe the neutrino position vector
${\bf r}=(r, \phi)$ where $r$ is the radius,  $\phi \in [0;2 \pi]$ is the polar angle, as shown in Fig.~\ref{bulb}.

The neutrino velocity can be decomposed in the radial ($v_r$) and transverse ($v_t$) component defined as 
${\bf v} = (v_r, v_{\pm}) = (\cos \theta_r, \pm \sin \theta_r)$, where
$\theta_r \in [0, \pi/2]$ is the angle between the radial direction and the one  of the neutrino propagation
(see, e.g.,~\cite{Buras:2005rp}),  and the $\pm$ sign indicate a transverse velocity in the clock-wise ($v_+$) or anti-clockwise ($v_-$) direction with respect to the radial direction, respectively.
We mention that in the recent multi-angle study~\cite{Abbar:2015mca}, where neutrinos emitted from a plane were considered, the range 
in the emission angles was $\theta \in [0, \pi]$. With this choice it is not necessary to distinguish  clock-wise or anti-clockwise modes. 
However, in our work we preferred to use have the $\theta_r \in [0, \pi/2]$ in order to start with a situation symmetric in the two emission directions $\pm$
and show the effect of breaking of this discrete symmetry.
Note that the local angle $\theta_r$ would depend on the radius $r$. 
In order to avoid this effect, in the literature it is  preferred to label the neutrino modes in terms of their emission angle
$\vartheta_{R} \in [0, \pi/2]$
along
the boundary at $r=R$. 
The two angles $\theta_r$ and $\theta_{R}$ are related by~\cite{EstebanPretel:2007ec}
\begin{equation}
R \sin \theta_R = r \sin \theta_r \,\ .
\label{eq:geom}
\end{equation}
Furthermore, we introduce the angular variable 
$u=\sin \theta_R$, $u \in[0,1]$.
With this choice the components of the neutrino velocity  are~\cite{Raffelt:2013rqa}
\begin{eqnarray}
v_r & = & \cos \theta_r = \sqrt{1-\frac{R^2}{r^2} u^2} \,\ , \nonumber \\
v_t \equiv v_{\pm} &=& \pm \sin \theta_r   = \pm \frac{R}{r} u  \,\ .
\end{eqnarray}

We assume that the neutrino distributions at $r=R$ in the energy $E$ and in the angular
variables $u$ and in the $R,L$ directions  can be factorized as
\begin{equation}
F_{\pm}(E, u) = F_{\nu}(E) \times F_{\nu}(u) \times F_{\pm} \,\ .
\end{equation}
We assume
 the neutrino angular distributions to be flat in $u$
and equal for all the flavors, i.e. $F_{\nu}(u)=1$.  
 
Concerning the  distributions in the clock-wise ($+$) or anti-clockwise ($-$) directions, we assume 
that these are given by
\begin{equation}
F_{\pm} = \frac{(1+\beta_{\pm})}{2+\beta_+ +\beta_-}  \,\ ,
\label{eq:anguldistr}
\end{equation}
where the quantities $\beta_{\pm} \ll 1$ are introduced to slightly perturb
the   $\pm$ of a given $u$ mode  at the boundary.

The neutrino number flux $F_{\nu}(E)$ 
at the  ring
 is given by
\begin{equation}
F_{\nu}(E) = \frac{1}{4 \pi R^2} \frac{L_{\nu}}{\langle E_\nu \rangle}  f_{\nu}(E) \,\ ,
\end{equation}
where we have normalized  the neutrino emission   
on a sphere with radius $R$.

\subsection{Two-flavor case}
In the following we will consider only a two-flavor system
$(\nu_e,\nu_x)$ where $x=\mu, \tau$ and we will describe 
the neutrino energy modes in terms of the two neutrino frequency
$\omega= \Delta m^2/2E_0$, where $\Delta m^2 = m_2^2 - m_1^2$ is the mass-squared difference. We have assumed a monochromatic
neutrino emission with $E=E_0$. In the two-flavor case 
the density matrices  are projected over the Pauli matrices $\sigma$  obtaining the  polarization vectors
in the usual way~\cite{EstebanPretel:2007ec},
where we normalize the (anti)neutrino polarization vectors to the difference of the anti-neutrino
fluxes at the boundary.

The Liouville operator on the left-hand-side of the EoMs  [Eq.~(\ref{eq:eom})] 
assumes the form
\begin{equation}
{\bf v} \cdot \nabla_{\bf r} = v_r \frac{d}{dr} + \frac{v_{\pm}}{r} \frac{d}{d\phi} \,\ ,
\end{equation}
so that the  EoMs  read (see also~\cite{Raffelt:2013rqa})
\begin{eqnarray}
\frac{d}{dr} {\sf P}_{\pm, u} &=& - \frac{v_{\pm}}{v_r r}\frac{d}{d\phi}{\sf P}_{\pm, u} \nonumber \\
&+& \left[\frac{\omega}{v_r} {\sf B} + \Omega^{\nu\nu}_{\pm} \right] \times {\sf P}_{\pm, u} \,\ ,
\label{eq:eompol}
\end{eqnarray}
where we indicated with sans-serif the vectors in flavor space. In particular, the unit vector ${\sf B}=({\sf B}^1, {\sf B}^2, {\sf B}^3)$ points  in the mass eigenstate direction in flavor space, such 
that ${\sf B}\cdot{\sf e}_3=-\cos \vartheta$, where $\vartheta$ is the vacuum mixing angle.
For simplicity we neglect a possible matter effect, assuming that its only role would be to reduce
the effective in-medium mixing angle, $\vartheta \ll 1$~\cite{Hannestad:2006nj}.  
The neutrino-neutrino interaction terms has a multi-angle kernel  $(1-{\bf v}_{\bf p} \cdot {\bf v}_{\bf q})$ which takes the form
\begin{eqnarray}
& &\frac{1}{v_r}\int d{\theta^\prime}_r [1-v_r {v_r^\prime} -v_t {v_t^\prime}] {\sf D^\prime} \nonumber \\
&=&\frac{R}{r} \int d{\vartheta_R^\prime} \cos{\vartheta_R^\prime} \left[\frac{1}{v_r {v_r^\prime}} -1 
- \frac{v_t {v_t^\prime}}{v_r {v_r^\prime}} \right] {\sf D^\prime}
\label{eq:multian}
\end{eqnarray}
where
${\sf D}$ is the difference between the neutrino and anti-neutrino polarization vector of a given mode, 
and  we used from Eq.~(\ref{eq:geom}) 
\begin{equation}
d \theta_r =\frac{R}{r} \frac{\cos{\vartheta}_R}{\cos \theta_r} d{\vartheta}_R
=\frac{R}{r}  \frac{\cos{\vartheta}_R}{v_r} d{\vartheta}_R = \frac{R}{r}  \frac{du}{v_r} \,\ .
\end{equation}
In the large-distance limit $(r \gg R)$
one can expand Eq.~(\ref{eq:multian}) obtaining
\begin{equation}
\frac{1}{2} \frac{R}{r}\int d{u^\prime} 
[v_t -{v_t^\prime}]^2 {\sf D^\prime}\,\ .
\end{equation}
We note that for the case we are studying the self-interaction term declines as $r^{-3}$, while 
in the SN case it declines as $r^{-4}$.
Considering the contribution of the clockwise ($+$)  and anti-clockwise ($-$) modes in the previous equation one gets
\begin{eqnarray}
{v_t}^2 {\sf D^\prime }&=&   \left(\frac{R}{r} \right)^2 u^2 ({\sf D}_{+, u^\prime}
+{\sf D}_{-, u^\prime}) \,\ \nonumber \\
v_t {v_t^\prime} {\sf D^\prime}  &=& \mp  \left(\frac{R}{r} \right)^2  {u u^\prime} 
({\sf D}_{+, u^\prime}
-{\sf D}_{-, u^\prime}) \,\ .
\end{eqnarray}
Then,   the neutrino self-interaction term 
in the large distance limit $r \gg R$ assumes the form
\begin{eqnarray}
\Omega^{\nu\nu}_{\pm} &=& \mu_r
 \int_{0}^{1} d u^\prime \large[(u^2 + {u^\prime}^2) \frac{({\sf D}_{+, u^\prime}
+{\sf D}_{-, u^\prime})}{2} \nonumber \\
&\mp & {u u^\prime}
({\sf D}_{+, u^\prime} -{\sf D}_{-, u^\prime}) \large]
 \label{eq:mularge}
 \end{eqnarray}
where  the $\mp$ refers to the  $\pm$  modes respectively, and
\begin{eqnarray}
\mu_r &=& [F_{{\bar\nu}_e}(R)- F_{{\bar\nu}_x}(R)] \frac{R^3}{2 r^3} \nonumber \\
&=& {3.5 \times 10^{5}} \,\ \textrm{km}^{-1}  \left(\frac{R}{r} \right)^3 \left(\frac{L_{\bar\nu_e}}{\langle E_{\bar\nu_e} \rangle}
-\frac{L_{\bar\nu_x}}{\langle E_{\bar\nu_x} \rangle}
 \right) \nonumber \\
&\times & \frac{15 \,\ \textrm{MeV}}{10^{51} \,\ \textrm{MeV}/\textrm{s}}\left(\frac{10 \,\ \textrm{km}}{R} \right)^2
 \,\ .
 \label{eq:mupotent}
\end{eqnarray}
An equation analogous to Eq.~(\ref{eq:multian}) can be written for the anti-neutrinos.

One can define a conserved ``lepton current'' 
${\textrm L}^\mu =({\textrm L}_0, {\bf L})$ whose components are
(see also~\cite{Duan:2008fd})
\begin{eqnarray}
{\textrm L}_0 &=& \int_{0}^{1} d u^\prime  \frac{1}{2}({\sf D}_{+, u^\prime} +
{\sf D}_{-, u^\prime}) \cdot{\sf B} \,\ , \label{eq:lepton0} \\
{\textrm L}_r &=& \int_{0}^{1} d u^\prime {v_r}  \frac{1}{2}({\sf D}_{+, u^\prime} +
{\sf D}_{-, u^\prime}) \cdot{\sf B} \,\ , \\
{\textrm L}_t &=& \int_{0}^{1} d u^\prime |{v_t}|  \frac{1}{2}({\sf D}_{+, u^\prime} -
{\sf D}_{-, u^\prime}) \cdot{\sf B} \,\ ,
\label{eq:lepton}
\end{eqnarray}
where  ${\bf L}$ is a two-dimensional vector  $({\textrm L}_r,{\textrm L}_\theta)$, and 
${\sf D}_{R(L), u} \cdot{\sf B} \simeq {\sf D}^3_{R(L),  u}$.  
From Eq.~(\ref{eq:eom}) one realizes that the lepton current satisfies a   continuity equation
\begin{equation}
\partial_{0} {\textrm L}_0 + \nabla_{\bf r} \cdot {\bf L} =\nabla_{\bf r} \cdot {\bf L}=0 \,\ ,
\label{eq:leptcons}
\end{equation}
where first equality follows since $\partial_0 {\textrm L}_0 =0$ having we  assumed  a  stationary solution.
 Eq.~(\ref{eq:leptcons}) generalizes the lepton-number conservation law 
 of the one dimensional case~\cite{Hannestad:2006nj}.

\subsection{Equations of motion in Fourier space}

The differential operators in Eq.~(\ref{eq:eompol}) implies that the flavor evolution 
is characterized by  a partial differential equation problem in $r$ and $\phi$. In~\cite{Mangano:2014zda,Mirizzi:2015fva} (see also~\cite{Duan:2014gfa}) it has been
shown how it is possible to solve such a problem by Fourier transforming the equations of motion with respect to the coordinate
along which a perturbation is introduced.
We  assume a perturbation of the polarization vectors at $r=R$ with period $2 \pi$ in  $\phi$ so that
\begin{equation}
{\sf P}_{\pm,u}(R, \phi) = {\sf P}^{0}_{\pm} +2 {\sf e }_z \delta \cos\phi  \,\ ,
\label{eq:polarseed}
\end{equation}
where ${\sf P}^{0}_{\pm}$ is the unperturbed value of the polarization vector, and $\delta \ll 1$ is the amplitude of the perturbation.
Up to the small difference in the emission modes  $\pm$  
[see Eq.~(\ref{eq:anguldistr})] the initial values of the polarization vectors are 
\begin{eqnarray}
{\sf P}^{0}_{\pm} (\nu) & \simeq &  (1 + \alpha) {\sf e}_z \,\ , \\
{{{\sf P}}}^{0}_{\pm} (\bar\nu) & \simeq & {\sf e}_z \,\ ,
\end{eqnarray}
where the initial flavor asymmetry is  given  by 
\begin{equation}
\alpha = \frac{F_{\nu_e} - F_{\bar{\nu}_e}}{F_{\bar{\nu}_e} - F_{\bar{\nu}_x}} \,\ .
\label{eq:alpha}
\end{equation}

The functions ${\sf P}_{\pm, u}(r, \phi)$ are periodic in $\phi$ with period $2\pi$. 
Their Fourier transform is then
\begin{equation}
{\sf P}_{\pm, u,n}(r) = \frac{1}{2 \pi} \int_{0}^{2\pi} {\sf P}_{\pm, u}(r, \phi) e^{-i n \phi} d \phi \,\ ,
\end{equation}
so that
\begin{equation}
 {\sf P}_{\pm, u}(r, \phi)  =\sum_{n={-\infty}}^{+\infty} {\sf P}_{\pm, u,n}(r) e^{+i n \phi} \,\ .
 \label{eq:invfour}
\end{equation}

The EoMs for the Fourier modes at large $r \gg R$ assume the form
\begin{eqnarray}
 & &\frac{d}{dr} {{{\sf P}}_{\pm, u,n}}(r) = \mp i n u \,\ \frac{R}{r^2} {{{\sf P}}_{\pm,u,n}} \nonumber \\
&+& \frac{\omega}{v_r} {\sf B}\times 
{{{\sf P}}_{\pm, u,n}}  \nonumber \\
&+& \mu_r   \sum_{j=-\infty}^{+\infty}  \int_{0}^{1}  d u^{\prime} 
[(u^2 + {u^\prime}^2)\frac{({{{\sf D}}_{+, u^\prime,{n-j}}}+ {{{\sf D}}_{-, u^\prime,{n-j}}})}{2} \nonumber \\
& &
 \mp {u u^\prime}
({{{\sf D}}_{+, u^\prime,{n-j}}} 
- {{{\sf D}}_{-, u^\prime,{n-j}}})
] \times {{{\sf P}}_{\pm, u,j}} \nonumber \,\ .
\label{eq:eompertradialb}
\end{eqnarray}
We stress that 
it is enough to follow the evolution for  positive modes $n \geq 0$, since the ${\sf P}_{\pm, u}(r, \phi)$ are real functions and therefore
\begin{equation}
{{\sf P}^{\ast}_{\pm, u,n}}= {{\sf P}}_{\pm, u,-n} \,\ .
\end{equation}
Once the evolution of the harmonic modes is obtained from Eq.~(\ref{eq:eompertradialb}), the polarization vector in configuration space can be obtained by inverse Fourier transform
[Eq.~(\ref{eq:invfour})].

\section{Numerical examples}

We present  the results of the flavor evolution in the two-dimensional model described above.
To calculate the $\nu$-$\nu$ interaction strength in Eq.~(\ref{eq:mupotent})
and the flavor asymmetry parameter in Eq.~(\ref{eq:alpha}) we use  benchmark values often used in previous studies of self-induced neutrino
oscillations (see, e.g.~\cite{Mirizzi:2010uz}), i.e. we take as average energies
\begin{equation}
(\langle E_{\nu_e} \rangle, \langle E_{{\bar\nu}_e} \rangle,
\langle E_{\nu_x} \rangle) = (12, 15, 18) \,\ \textrm{MeV} \,\ ,
\end{equation}
while for the neutrino luminosities (in units of $10^{51}$~erg/s) we assume
\begin{equation}
L_{\nu_e}=2.40 \,\ \,\  , \,\ \,\  L_{\bar\nu_e}=2.0 \,\ \,\  , \,\ \,\   L_{\nu_x}=1.50 \,\ .
\end{equation}
These values are typical of the early time SN accretion phase, and corresponds to an asymmetry parameter
$\alpha=1.34$. 
As specified before we work in a single-energy scheme, where we take as representative vacuum oscillation
frequency
the one corresponding to the average of the $\nu$ ensemble with the emissivity parameters  
chosen above (see~\cite{Mirizzi:2010uz}).
Namely we take $\omega= 0.68$~km$^{-1}$.
Concerning the neutrino oscillation parameters we choose a small mixing angle $\vartheta=10^{-2}$.
Moreover, we assume $N_u=100$ modes for the angular variable $u \in [0;1]$ in order to have numerical convergence of the results
and to  avoid spurious instabilities
due to few  angular modes.

\begin{figure}[!t]\centering
\includegraphics[angle=0,width=1.\columnwidth]{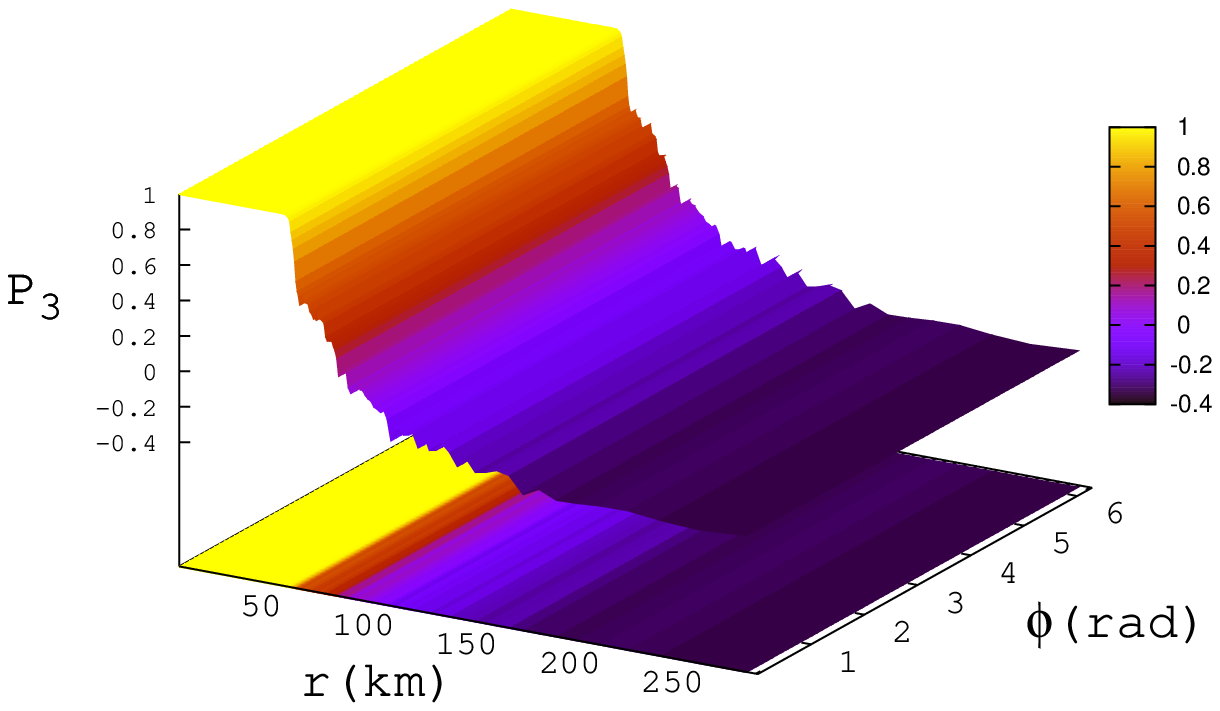}
\vspace{2.cm}
\includegraphics[angle=0,width=1.\columnwidth]{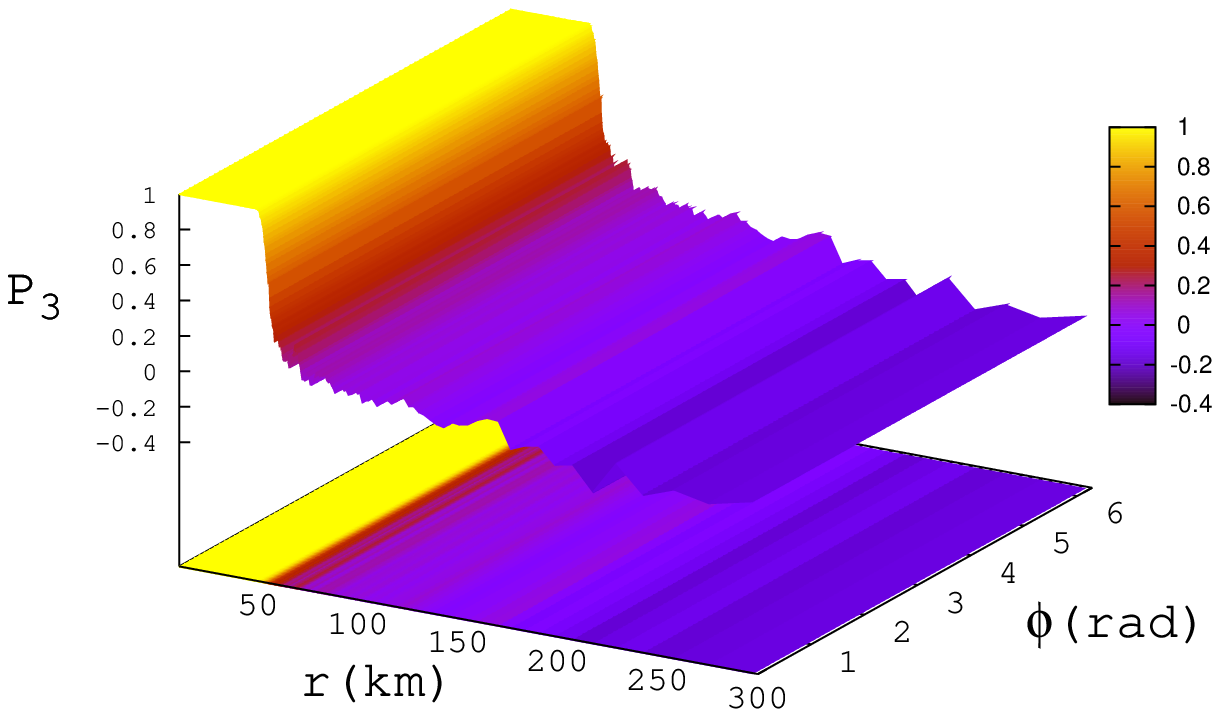}
\vspace{-2.cm}
\caption{Two-dimensional evolution of the $3$-rd component $P_3$ of the  ${\bar\nu}$ polarization vector  in the $r$-$\phi$ plane, and its map  on the bottom plane
breaking only the $\pm$  symmetry. Upper plot refers to NH, while lower panel is for IH. 
\label{pzt}}
\end{figure}

\begin{figure}[!t]\centering
\includegraphics[angle=0,width=1.\columnwidth]{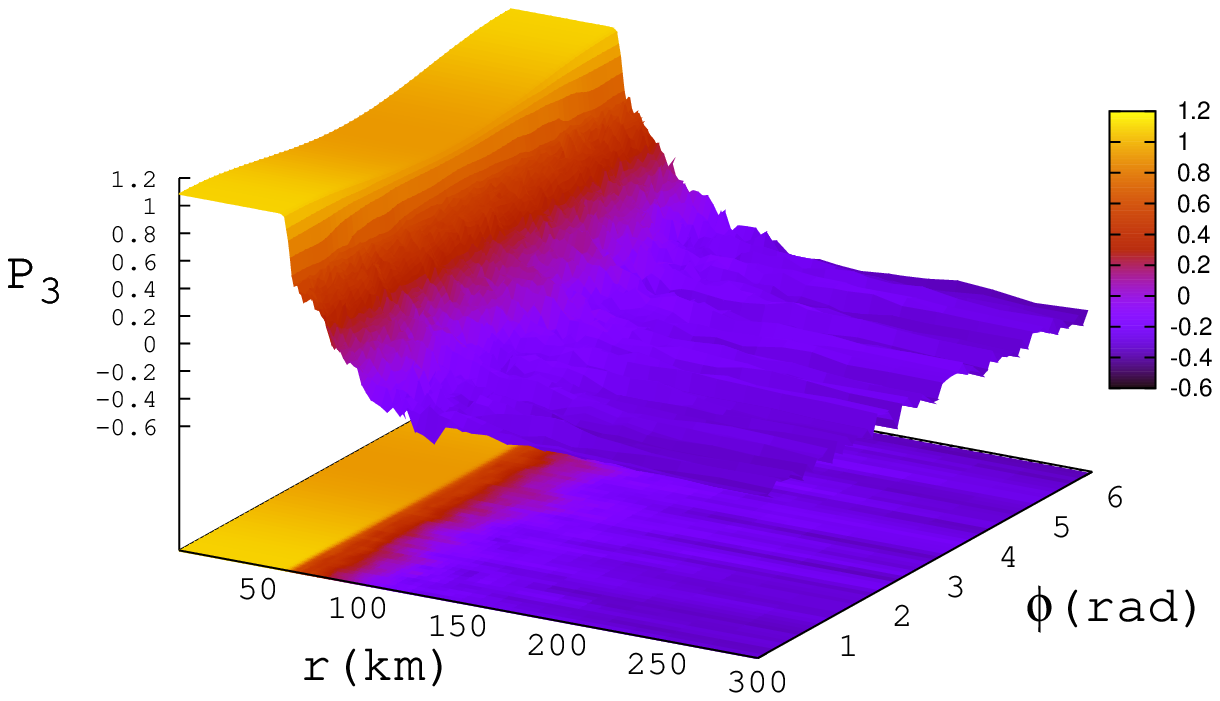}
\vspace{2.cm}
\includegraphics[angle=0,width=1.\columnwidth]{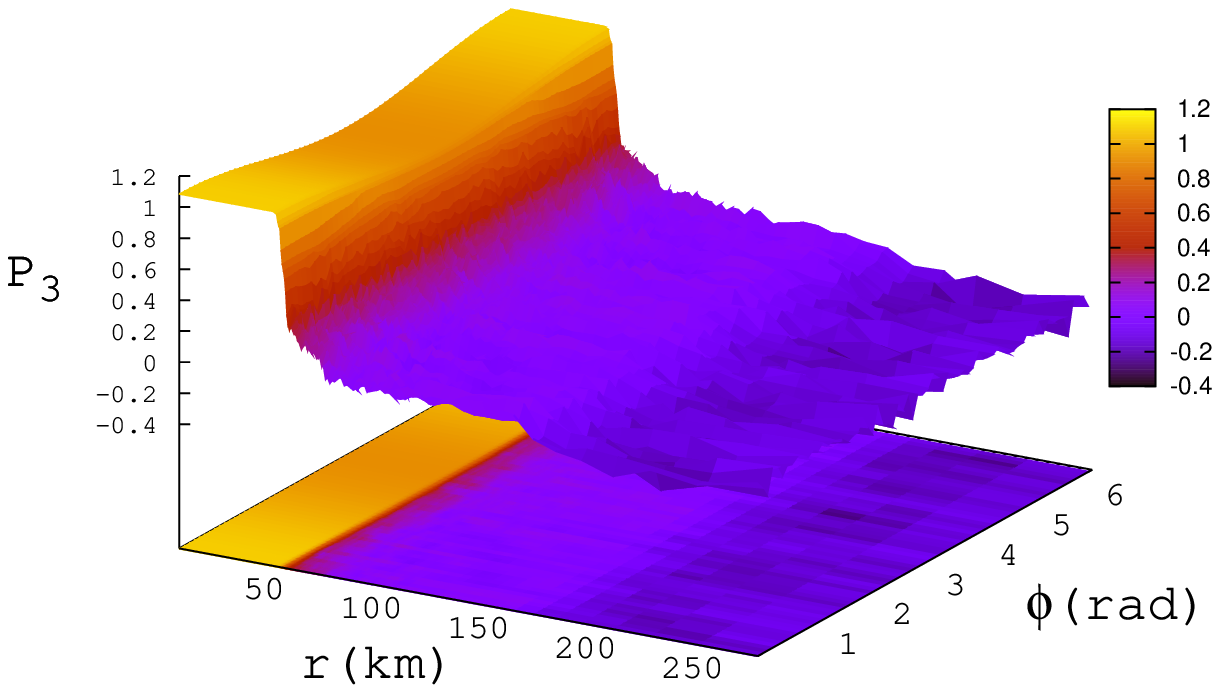}
\vspace{-2.cm}
\caption{Two-dimensional evolution of the $3$-rd component $P_3$ of the  ${\bar\nu}$ polarization vector  in the $r$-$\phi$ plane, and its map  on the bottom plane
breaking the azimuthal  invariance for   the $\pm$  modes and the translational symmetry on the ring. Upper plot refers to NH, while lower panel is for IH. 
\label{pznot}}
\end{figure}

It is known   that  forcing the  $\pm$ symmetry (taking $\beta_+=\beta_-=0$ in Eq.~(\ref{eq:anguldistr}))
and the translational symmetry on the ring (taking $\delta=0$ in Eq.~(\ref{eq:polarseed}))  the ensemble is stable in normal
mass hierarchy (NH, $\Delta m^2 >0$) while in inverted mass hierarchy (IH, $\Delta m^2 <0$) it exhibits
large  bimodal flavor changes in the form of \emph{pair conversions}
$\nu_e \bar\nu_e \to \nu_x\bar\nu_x$~\cite{Hannestad:2006nj}. 
If we perturb the $\pm$ symmetry taking small seeds $\beta_+ = -\beta_-$ in the distributions
of  Eq.~(\ref{eq:anguldistr}) the system now exhibits the analogous of  the so-called    \emph{multi-azimuthal-angle} (MAA) 
instability of the bulb model~\cite{Raffelt:2013rqa,Mirizzi:2013rla,Mirizzi:2013wda}. 
The result of the flavor evolution is shown in Fig.~\ref{pzt} where it is represented the behavior of the $3$-rd component
of the (integrated over $u$) anti-neutrino polarization vector ${\sf P}_3 = 1/2 ({\sf P_+}+{\sf P_-})$ in the $(r, \phi)$ plane for NH (upper panel) and IH (lower panel).
In these numerical 
examples we have chosen $\beta_+= 10^{-2}$. 
 The most striking effect of the MAA instability is that now also NH exhibits flavor conversions at $r \gtrsim 60$~km. 
The choice of the initial seed $\beta_{\pm}$ determines the onset radius of the flavor conversions: the largest 
the seed, the earliest flavor conversions start.
 In IH (lower panel) flavor conversions start as in the $\pm$ symmetric case at $r \gtrsim 50$~km and MAA 
 has a minor impact on the flavor evolution.
From the Figure we realize that the behavior of the flavor conversions is uniform in the $\phi$ variable since the translational symmetry
has remained unbroken. Indeed we have solved only the EoMs [Eq.~(\ref{eq:eompertradialb})] for the $n=0$ Fourier mode in this case.

\begin{figure}[!t]\centering
\includegraphics[angle=0,width=0.8\columnwidth]{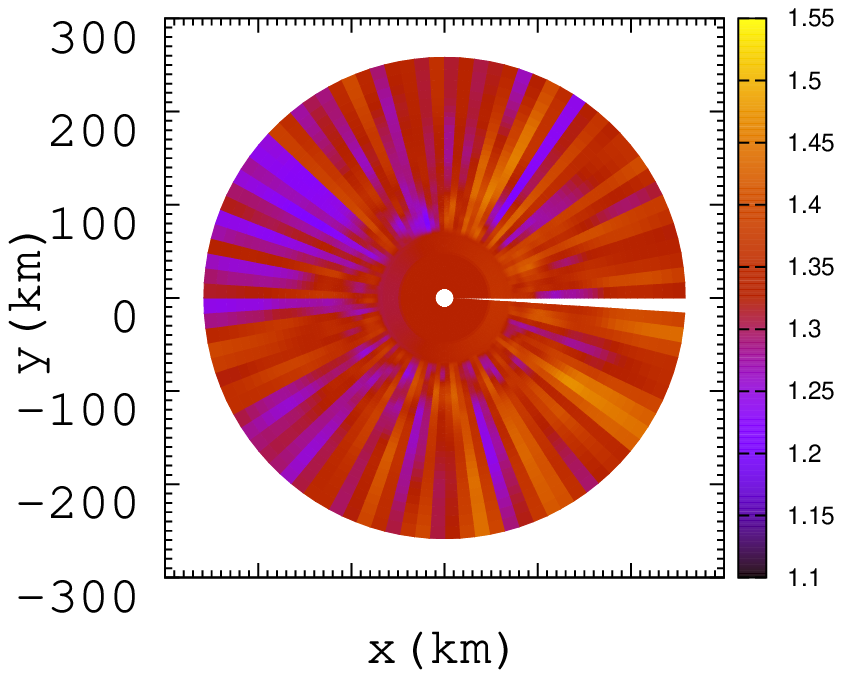}
\vspace{2.cm}
\includegraphics[angle=0,width=0.8\columnwidth]{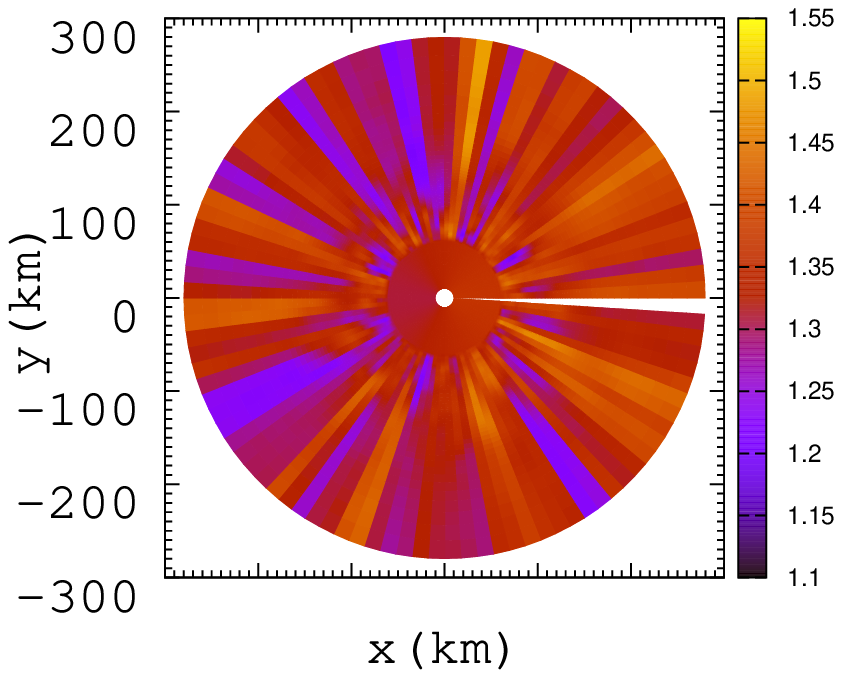}
\vspace{-2.cm}
\caption{Component ${\textrm L}_r$ of the vector lepton number ${\bf L}$ in cartesian coordinates in NH (upper panel)
and IH (lower panel), respectively. 
\label{leptocart}}
\end{figure}

The next step is to perturb also the translational symmetry on the ring assuming a seed $\delta$ in the 
 longitutudinal distribution of the polarization vectors on the boundary [see Eq.~(\ref{eq:polarseed})].
 In this case we consider the evoltion of the first N=100 Fourier modes in Eq.~(\ref{eq:eompertradialb}).
 In this way we are sensitive to variations occuring  at an angular  scale $\Delta \phi \gtrsim 3^{\circ}$. 
  Results are shown in Fig.~\ref{pznot} with the same format of the previous Figure.
 We used as seed to break the $\phi$-symmetry
$\delta = 3 \times 10^{-3}$.
In both NH  (upper panel) and IH case (lower panel) flavor conversions start as in 
the translational invariant case,
 i.e.
the planes of common oscillation phase are flat in $\phi$ direction.
 However this behavior is not stable.
In the NH case 
around
$r \simeq 100$~km something  occurs:
The ${\sf P}_3$ component is no longer flat in $\phi$, while it 
starts to acquire notable variations ($\sim 20~\%$) at different  longitude.
In IH  after  flavor conversions develop, also the translational symmetry is perturbed at $r \gtrsim 100$~km, 
with variations  at different $\phi$ with values up to  $30~\%$.

In Fig.~\ref{leptocart} we represent the component ${\textrm L}_r$ of the vector lepton number ${\bf L}$
 [Eq.~(\ref{eq:lepton})]
in cartesian coordinates
\begin{eqnarray}
x &=& r \cos \phi \,\ , \nonumber \\
y &=& r \sin \phi \,\ .
\end{eqnarray}
We realize that when the spherical symmetry is broken, the lepton number 
acquires significant variations in different directions at a given $r$ with respect to the initial uniform value
${\textrm L}_r=\alpha=1.34$. 
In particular 
 one  finds  $\sim 20~\%$ variations.

\begin{figure}[!t]\centering
\includegraphics[angle=0,width=1.\columnwidth]{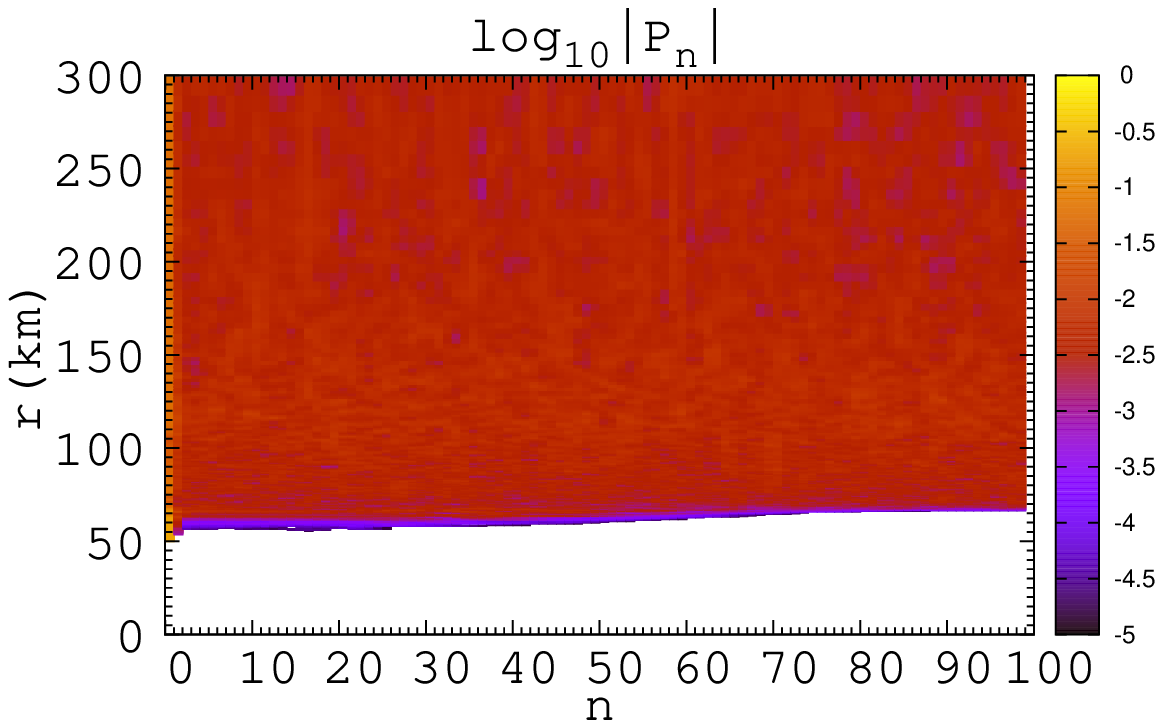}
\vspace{2.5cm}
\includegraphics[angle=0,width=1.\columnwidth]{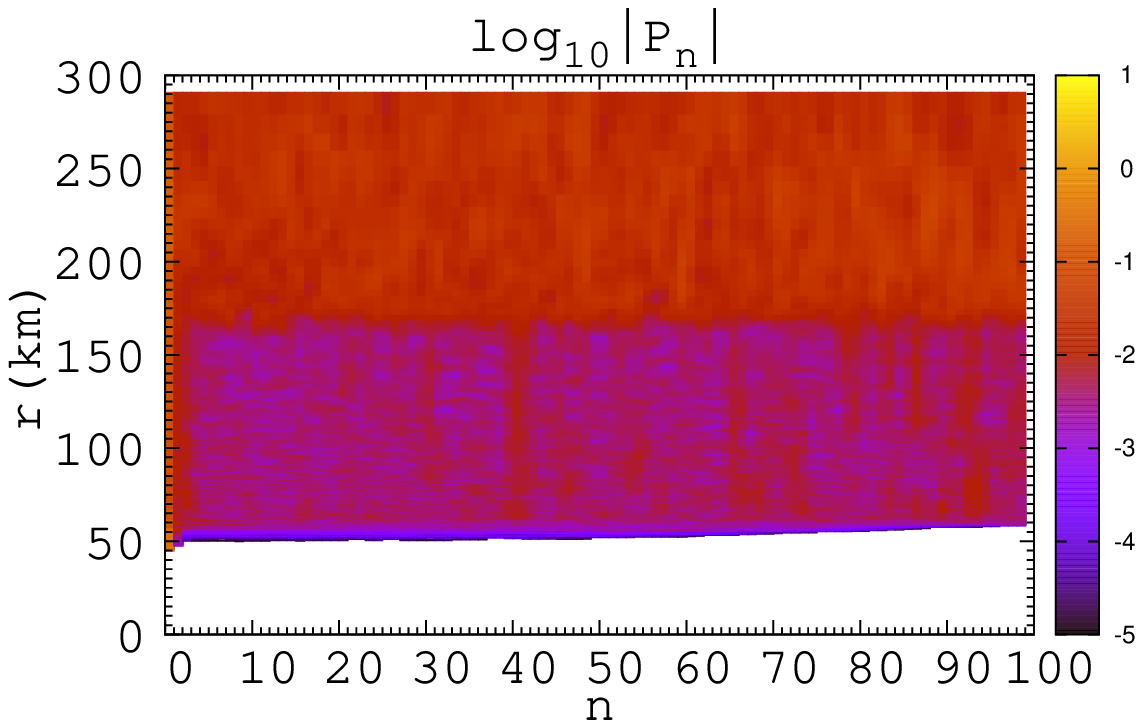}
\vspace{-3.cm}
\caption{Contour plots of the first 100 Fourier modes $|{\sf P}_{n}|$ (in logarithmic scale) in the plane $n$-$r$
in NH (upper panel) and IH (lower panel) respectively.
\label{normcart}}
\end{figure}

In order to clarify better this flavor dynamics, in Fig.~\ref{normcart} we show a contour plot representing 
the growth of the different Fourier modes $|{\sf P}_n|$ (in logarithmic scale) in the plane $n$-$r$
for NH (upper panel) and IH (lower panel). We consider the evolution of the first $N=100$ modes.
We realize that the breaking of the translational symmetry at $r \simeq 60$~km 
in NH corresponds to the rapid excitation of the $n >0$ harmonics that reach values
$|{\sf P}_n| \lesssim  10^{-2}$. 
Instead in IH for $r \gtrsim 50$~km the modes start to get excited and can grow to  $|{\sf P}_n| \gtrsim  10^{-1.5}$ 
at $r \gtrsim 150$~km. This explains 
why the effect of the  breaking of the translational symmetry is more 
pronounced in IH rather than in NH.

\section{Conclusions}

We have considered a simple two-dimensional toy-model, namely  neutrino beams emitted from a ring, to point-out the effect of spontaneous breaking of axial and 
spherical symmetries in the self-induced flavor conversions of SN neutrinos.

We found that if the slightly perturb the space symmetries on the boundary, 
these perturbation seeds are dramatically amplified  altering the flavor conversions found in a symmetric model.
Therefore, the flavor content of the self-interacting SN neutrinos would acquire significant direction-dependent variations.
These results are qualitatively similar to what we found in the planar model we studied in~\cite{Mirizzi:2010uz}.  
Our findings suggest that the characterization on the flavor conversions obtained before should be critically reconsidered,  including 
these spontaneous symmetry breaking effects.
In order to have a realistic characterization of the possible SN neutrino spectra our simple toy model should be improved on different aspects.
In particular, one should extend this model to a  realistic three-dimensional spherical case. 
In this case one would have the possibility to break the spherical symmetry in both longitudinal
and latitudinal directions.
Moreover, in order to get our numerical solution  we have considered $N=100$ Fourier modes. In this cases we have not found the presence of flavor converions
at lower radii than in the spherically symmetric case. However, in~\cite{Duan:2014gfa} it has been shown that  harmonics with sufficiently high $n$ could
become unstable also at low-radii. Increasing $N$ to 500 we have not found any sizeble change in the onset of the flavor changes and in the subsequent flavor evolution
in the non-linear regime. However, it remains 
to be seen if with a  much higher number of harmonics low-radii effects could occur. 
In this regard, it would be useful a stability analysis performed along the lines of~\cite{Chakraborty:2015tfa}.

Continuous energy spectra should also be taken into account to understand how the spectral splitting features found in the bulb model would be modified in this case.
The role of matter effects that would suppress self-induced flavor conversions during the accretion phase should also be investigated.
The final goal would be to  study of the self-induced neutrino flavor conversions in
realistic multi-dimensional supernova models accounting for largely
 aspherical neutrino emission and matter   profiles.
 This objective
is particularly timely now since in the last recent years, SN model simulations have
experienced several breakthroughs. After 1D~\cite{Fischer:2009af} and 2D~\cite{Buras:2005rp} models, the forefront has reached 3D
SN simulations~\cite{Wongwathanarat:2014yda,Lentz}. Therefore, it seems the perfect juncture to connect realistic SN simulations
with nonlinear neutrino oscillations.
This open issue makes compulsory the need for further dedicated studies to fully clarify the fascinating behavior of the interacting neutrino field.

\section*{Acknowledgements} 

The author warmly thanks Pasquale Serpico for useful comments on this manuscript.
This work is supported by the Italian Ministero dell'Istruzione, Universit\`a e Ricerca (MIUR) and Istituto Nazionale
di Fisica Nucleare (INFN) through the ``Theoretical Astroparticle Physics'' projects.

\end{document}